\documentclass[12pt,a4,epsfig]{article}
\usepackage{graphicx}
\newcommand{\gb}{Gauss-Bonnet term}
\newcommand{\be}{\begin{equation}}
\newcommand{\ee}{\end{equation}}
\newcommand{\bc}{\begin{center}}
\newcommand{\ec}{\end{center}}

\newcommand{\integrat}{\int d^5x \sqrt{-g}}

\newcommand{\lb}{\left(}
\newcommand{\rb}{\right)}

\newcommand{\asch}{AdS-Schwarzschild}
\newcommand{\sectiono}[1]{\section{#1}\setcounter{equation}{0}}
\newcommand{\p}{\partial}
\newcommand{\ben}{\begin{eqnarray}\displaystyle}
\newcommand{\een}{\end{eqnarray}}

\newcommand{\ra}{\rightarrow}

\newcommand{\sh}{Schwarzschild}
\newcommand{\gc}{ grand-canonical ensemble }

\newcommand{\cmp}{ chemical potential $\Phi$ }
\newcommand{\adsrn}{AdS-Reissner-Nordstr\" om black hole}
\newcommand{\adsrns}{AdS-Reissner-Nordstr\" om black holes}
\newcommand{\rn}{AdS-Reissner-Nordstr\" om}

\begin{document}
\noindent

{}~
{}~

\hfill\vbox{\hbox{0705.2682[hep-th]}}\break

\vskip .6cm

{\baselineskip20pt
\begin{center}
{\LARGE {\bf   
Phase Transition of Electrically Charged Ricci-flat Black Holes
}}
\end{center} }
\begin{center}
{\bf \underline{ \ \ \ \ \ \ \ \ \ \hskip 13cm }}
\end{center}
\vskip .6cm
\medskip

\vspace*{4.0ex}

\centerline{ {\bf \large Nabamita Banerjee} and {\bf \large Suvankar
    Dutta} }

\vspace*{4.0ex}

\centerline{\large \it Harish-Chandra Research Institute}

\centerline{\large \it  Chhatnag Road, Jhusi,
Allahabad 211019, INDIA}

\vspace*{1.0ex}

\centerline{E-mail: {\it nabamita@mri.ernet.in,suvankar@mri.ernet.in}}

\vspace*{5.0ex}

\centerline{\bf Abstract}
\bigskip

We study phase transition between electrically charged Ricci-flat
black holes and AdS soliton spacetime of Horowitz and Myers in five
dimensions.  Boundary topology for both of them is ${\bf S^1 \times
S^1 \times R^2}$.  We consider \rn \ black hole and R-charged black
holes and find that phase transition of these black
holes to AdS soliton spacetime depends on the relative size of two
boundary circles.  We also perform the stability analysis for these
black holes. In order to use the AdS/CFT correspondence, we work in
the grand canonical ensemble.

\vfill \eject

\baselineskip=18pt

{}

\tableofcontents


\sectiono{{\bf Introduction}}

The correspondence between supergravity in anti-de Sitter(AdS)
spacetime and conformal field theory in one lower dimension is one of
the interesting subjects of current research. It has been conjectured
\cite{maldacena} by Maldacena that the type IIB supergravity
(superstring theory) on ${\bf AdS_5 \times S^5}$ is dual to the the
$\cal {N}$ = 4 super Yang-Mills theory living on the boundary of the
AdS space. The conjecture also relates thermodynamics of the gauge
theory on conformal boundary of the AdS space with that of the gravity
residing in the bulk AdS.

We will first review thermodynamics of the gauge theory on a three
sphere ${\bf S^3}$. To compute the free energy or entropy of the
finite temperature gauge theory, we have to first calculate the
partition function on ${\bf S^3 \times S^1}$, where ${\bf S^1}$ is the
Euclidean time circle. The circumference $\beta$ of ${\bf S^1}$ is
related to the inverse temperature of the field theory and we denote
the radius of ${\bf S^3}$ as $\beta '$.  Due to the conformal
invariance of the gauge theory on ${\bf S^3 \times S^1}$, only the
ratio of these two parameters $\beta/\beta'$ is relevant.

Phase transition is one of the important aspects in the study of
thermodynamics.  A system in the finite volume, in general, does not
exhibit any phase transition. But in the large $N$ limit, {\it i.e.},
when the number of degrees of freedom goes to infinity then it is
possible to have a phase transition even in finite volume
\cite{grosswitten}.  It has been argued in \cite{witten1} that the
$\cal{N}$ =4 SYM theory on ${\bf S^3 \times S^1}$ shows a phase
transition as a function of $\beta /\beta '$ in large $N$ limit. The
large $\beta /\beta '$ or small temperature phase corresponds to the
confining phase where small $\beta /\beta '$, {\it i.e.}, the large
temperature phase corresponds to de-confining phase of the gauge
theory.

On the gravity side Hawking and Page had shown in \cite{hawking} that
there exists a phase transition (HP phase transition) between
spherical AdS (\sh) black hole and and global AdS spacetime.  Above
the transition temperature (Hawking-Page transition temperature), the
black hole spacetime is stable while below that temperature the black
hole spacetime is unstable and decays to the global AdS space time.

Using the AdS/CFT correspondence, Witten has identified the
 Hawking-Page phase transition in gravity side with the
 confinement-deconfinement phase transition in the gauge theory
 \cite{witten1,witten2}.  The low temperature phase on the gravity
 side which is dominated by global AdS spacetime corresponds to the
 confining phase of gauge theory while the high temperature phase
 where AdS black hole is energetically favored corresponds to the
 deconfining phase of the gauge theory.

The phase transition, both on the gravity side and on the gauge theory
side, is sensitive to the topology of spacetime.  For example, instead
of spherical asymptotic geometry if one considers AdS black holes with
planar asymptotic geometry then it is easy to show that there exists
no HP transition between this black hole phase and the global AdS
spacetime.  In other words, the planar black hole phase is always
dominant for any non zero temperature.  On the other hand dual thermal
gauge theory of this planar black hole, which is defined on ${\bf S^1
\times R^3}$ also does not show any phase transition.  The gauge
theory is always in deconfined phase. This can be understood from the
fact that planar geometry can be obtained from spherical geometry in
$\beta' \ra \infty$ limit and in this limit $\beta /\beta' \ra 0$.
The obvious question to ask here is, what happens when the black hole
spacetime has asymptotic topology other than ${\bf S^1 \times S^3}$ or
${\bf S^1 \times R^3}$?  We will try to address this question in the
context of charged black holes in this paper.

To answer this question we have to first find out what kind of black
hole topologies one can have in asymptotically AdS space. In fact, it
is possible to construct a black hole solution of Einstein equation
with positive (spherical), zero (flat) or negative (hyperboloid)
constant curvature horizons in AdS spacetime \cite {cai er paper}.
Due to different horizon topologies, the thermodynamic properties of
these black holes are different.  We will be focusing on black holes
with flat horizon.  It has been shown in \cite{vanzo, birmi} that
there are no phase transitions between the AdS black hole with Ricci
flat horizon and the zero mass black hole (global AdS spacetime)
\footnote{It is worth mentioning here that in studying the
  thermodynamics of Euclidean black hole a proper background
  subtraction is essential to get finite thermodynamic variables.  One
  also has to ensure that the boundary topology of black hole
  spacetime and that of the background spacetime are same
  asymptotically (see \cite{dutta} for an explicit discussion).  For
  example, in asymptotically AdS space if we want to study the
  thermodynamics of \asch black hole then the background we choose is
  asymptotically pure AdS spacetime.} \footnote{see \cite{myung} for
  related discussion in de-Sitter space and \cite{papa} for
  topological black holes with hair.}.

However, in \cite{sumati1}, the authors have shown that it is possible
to find a phase transition between Ricci-flat black holes and a AdS
soliton spacetime where both the spacetime has asymptotic geometry is
${\bf S^1 \times S^1 \times R^2}$ ( ${\bf S^1\times R^2}$ is the
Ricci-flat space) \footnote{Higher derivative correction to the phase
  transition of Ricci flat neutral black hole has been studied in
  \cite{nojiri} and in a recent paper \cite{cai0}.}.  It has been
conjectured by Horowitz and Myers \cite{horowitz-myers} that given the
Ricci-flat boundary topology ${\bf S^1 \times S^1 \times R^2}$ the AdS
soliton spacetime is the minimum energy (perturbatively stable)
solution of the Einstein equation. We will review briefly this
conjecture of Horowitz and Myers and the phase transition between
neutral black holes and AdS soliton spacetime in section \ref{review}.

Our main focus in this paper will be to generalize this analysis to
charged black holes.  Thermodynamics of charged black holes typically
has more interesting features than the neutral ones. Lower dimensional
charged black holes in the AdS space can arise in the 
compactification of string theory on ${\bf S^{n+1}}$.  One simple way
to get a charged black hole solution in lower dimensions (say, in 5
dimensions) is by compactifying the string theory (type IIB) on a
rotating sphere ({\it i.e.}, ${\bf S^5}$ in this
case)\cite{tenauth,jonson}.  In this paper, we will discuss the
thermodynamics of Ricci flat charged black holes and phase transition
between the black hole and AdS soliton spacetime.

Organization of our paper is as follows. In section \ref{review} we
briefly review the AdS soliton spacetime (section \ref{horo-myers})
and phase transition of neutral Ricci flat black holes (section
\ref{sumathi}).  In the next section (section \ref{chargebh}) we
discuss the thermodynamics of charged Ricci-flat black holes in the
grand canonical ensemble and study phase transition into AdS soliton
spacetime.  We show that the phase transition line depends on the size
of the spatial ${\bf S^1}$ circle. We also discuss the thermodynamic
stability of these black holes in section \ref{stab}.  The last
section \ref{dis} contains discussion of our results.  The detailed
calculation leading to the on-shell Euclidean action for general
R-charged black holes is given in the appendix \ref{appA}.



\sectiono{ {\bf AdS Soliton Spacetime and Phase Transition of
    Ricci-flat Black Holes}}\label{review}

In this section we briefly review the work of Horowitz and Myers
\cite{horowitz-myers} and discuss how asymptotically Ricci flat black
hole undergoes a phase transition into the AdS soliton spacetime
\cite{sumati1}.

\subsection{AdS Soliton Spacetime}\label{horo-myers}

The AdS soliton spacetime is given by the metric, 
\be\label{adssol1}
ds_S^2 = -{r^2 \over b^2} dt_S^2 + {b^2 \over r^2}\lb 1 - {\mu^4 \over
r^4} \rb^{-1} dr^2 + {r^2 \over b^2}\lb 1 - {\mu^4 \over r^4} \rb
d\theta_S^2 + {r^2 \over b^2} h_{ij}dx^idx^j, 
\ee 
where, $b$ is radius of the AdS spacetime, $\mu$ is a constant
parameter related to the energy of this spacetime and $h_{ij}$ is the
metric on a two dimensional Ricci-flat manifold $R^2/\Gamma$, where
$\Gamma$ is a finite discrete group.  This two dimensional manifold
${\bf Y^2}$ can be a torus ${\bf T^2}$ for some non-trivial $\Gamma$
or in the simpler case it is ${\bf R^2}$.  We will consider ${\bf
  Y^2}$ to be ${\bf R^2}$ throughout this paper.  This metric can be
obtained as a solution of the Einstein equation (which is obtained by
varying the Einstein-Hilbert action with negative cosmological
constant).  The metric can also be obtained by a double analytic
continuation of a five dimensional Ricci-flat AdS black hole metric
with $t \rightarrow i \theta_S$ and $\theta \rightarrow it_S$ (see
\cite{takayanagi} for related discussion),
\be
ds^2 = -{r^2 \over b^2}\lb 1 - {\mu^4 \over r^4} \rb dt^2 + {b^2 \over
r^2}\lb 1 - {\mu^4 \over r^4} \rb^{-1} dr^2 + {r^2 \over b^2}
d\theta^2 + {r^2 \over b^2} h_{ij}dx^idx^j.  \ee 
In (\ref{adssol1})
the coordinate $r$ is restricted to $r \geq \mu$ and $\theta_S$ must
be identified with period, 
\be \label{soltemp} \eta_S = { \pi b^2
\over \mu} \ee 
so as to avoid the conical singularity at $r=r_0$. The
soliton spacetime $X_S$ has a topology ${\bf R \times {\cal B}^2
\times Y^2}$, where ${\bf {\cal B}^2}$ is a two dimensional ball and
${\bf Y^2}$ is either ${\bf T^2}$ or ${\bf R^2}$ as mentioned above. 
Boundary $M_S$ of this soliton spacetime has the topology of ${\bf
R\times S^1 \times Y^2}$.

The AdS soliton spacetime has negative energy and it is given by, 
\be
\label{solen} E_S= - {V b^3 \pi^3 \over 16 \pi G_5 \eta_S^3}, 
\ee
where $V= \int \sqrt h d^2x$.  We can write this energy in terms of
gauge theory variables using the AdS/CFT dictionary,
\ben\label{dictionary}
b^4 = {N \sqrt{2 G_{10}} \over \pi^2}, \nonumber \\
G_5 = {G_{10} \over Vol(S^5)}
\een
as, 
\be\label{solenergy-dic}
E_S=-{V \pi^2 N^2 \over 8 \eta_S^3}.
\ee
The energy density of this spacetime then becomes,
\be\label{solenden}
\rho_{S} = {E \over V \eta_S} = - {\pi^2 N^2 \over 8 \eta_S^4}.
\ee

Using the AdS/CFT correspondence we can compare the energy density of
this soliton spacetime with the ground state energy of the dual field
theory on ${\bf S^1 \times Y^2}$ where the length of $S^1$ is ${\bf
  \eta_S}$.  Here the field theory is ${\cal N}$=4 SYM theory with
SU(N) gauge group.  The fermions are antiperiodic on ${\bf S^1}$.  To
compare the result obtained on the gravity side, we need to determine
the Casimir energy of the weekly coupled gauge theory on the boundary
of AdS soliton spacetime with the same boundary condition for fermions
in ${\bf S^1}$ direction. When ${\bf Y^2 = R^2}$, the leading order
result for the Casimir energy density is given by\footnote{The general
  case, {\it i.e.}, when the boundary topology is $R \times T^3$ or
  $R^2 \times T^2$ rather than $R^3 \times S^1$ has been studied in
  \cite{myers}.} \cite{casimir},
\be\label{casienden}
\rho_{gauge} = - {\pi^2 N^2 \over 6 \eta_S^4}.
\ee
Thus the negative energy density of the AdS soliton spacetime is
precisely 3/4 of the Casimir energy of the zero coupling gauge theory.
The discrepancy of a factor 3/4 reflects the fact that two results
apply in different regime of the dual gauge theory.

The above agreement is in favor of AdS/CFT correspondence in a
non-supersymmetric case.  But, the important question is whether the
AdS soliton solution (\ref{adssol1}) is the lowest energy stable
solution of the Einstein equations for the given boundary topology.
On the basis of the stability of the non-supersymmetric field theory
on ${\bf S^1 \times Y^2}$ together with the help of AdS/CFT
correspondence, Horowitz and Myers have conjectured that the AdS
soliton solution (\ref{adssol1}) is the minimum energy solution of the
Einstein equations with that boundary condition.  Any other solution
with this boundary condition has positive energy with respect to
(\ref{solen}).  They have also shown that the candidate minimum energy
solution is stable against all quadratic fluctuations of the metric.

\subsection{Phase Transition of Ricci-flat Black Hole} \label{sumathi}

We shall now review how Ricci-flat black holes undergo a phase
transition into the AdS soliton spacetime \cite{sumati1}.  The
Ricci-flat black hole metric is given by,
\be\label{adsbh} 
ds^2 = -{r^2 \over b^2}\lb 1 - {r_0^4 \over r^4} \rb
dt^2 + {b^2 \over r^2}\lb 1 - {r_0^4 \over r^4} \rb^{-1} dr^2 + {r^2
  \over b^2} d\theta^2 + {r^2 \over b^2} h_{ij}dx^idx^j, 
\ee
where $r_0$ is the position of the horizon and the periodic coordinate
$\theta$ has a period $\eta$. Both the black hole metric and the AdS
soliton metric are obtained varying the action,
\be\label{acn}
I = -{1 \over 16 \pi G_5} \int d^5 x \sqrt{-g} \lb R + {12 \over b^2}
\rb.
\ee

To study the thermodynamics of this black hole spacetime one has to go
to the Euclidean theory by a Wick rotation $t \rightarrow i \tau$. The
Euclidean metric is given by,
\be\label{euclibh} 
ds^2 = {r^2 \over b^2}\lb 1 - {r_0^4 \over r^4} \rb
d\tau^2 + {b^2 \over r^2}\lb 1 - {r_0^4 \over r^4} \rb^{-1} dr^2 +
{r^2 \over b^2} d\theta^2 + {r^2 \over b^2} h_{ij}dx^idx^j 
\ee
for the black hole spacetime and
\be\label{euclisol}
ds_S^2 = {r^2 \over b^2} d\tau_S^2 + {b^2 \over r^2}\lb 1 - {\mu^4 \over
  r^4} \rb^{-1} dr^2 + {r^2 \over b^2}\lb 1 - {\mu^4 \over r^4} \rb
d\theta_S^2 + {r^2 \over b^2} h_{ij}dx^idx^j
\ee
for the soliton spacetime. The regularity of black hole spacetime at
$r=r_0$ demands that the Euclidean time coordinate $\tau$ of the black
hole is identified with period,
\be\label{bhtemp}
\beta = {\pi b^2 \over r_0}, 
\ee
where the Euclidean time coordinate $\tau_S$ of the soliton space time
has an arbitrary period $\beta_S$. It is worthwhile to mention here
that $1/\beta$ and $1/\beta_S$ are the temperature of the black hole
and soliton spacetime respectively. 

We will now discuss the phase transition between the 
black hole spacetime and the AdS 
soliton spacetime. Given the boundary geometry ${\bf R \times S^1 \times R^2}$ 
there are three solutions to the Einstein equations derived from 
 the action (\ref{acn}). {\bf (a)} Black hole 
spacetime (\ref{adsbh}), {\bf (b)} soliton spacetime (\ref{adssol1}) and 
{\bf (c)} global AdS spacetime, whose metric is given by,
\be
ds_{AdS}^2 = -{r^2 \over b^2} dt_{AdS}^2 + {b^2 \over r^2} dr^2 + {r^2
  \over b^2} d\theta_{AdS}^2 + {r^2 \over b^2} h_{ij}dx^idx^j.
\ee
 If we compute the on-shell action or 
free energy for black hole spacetime with respect to global AdS spacetime 
we find no signature of phase transition, {\it i.e.}, the 
black hole spacetime is always
dominant over the global AdS spacetime. Similarly there exists no 
phase transition between the AdS soliton and the global AdS.  The black
hole free energy is given by,
\be\label{fb}
F_{B} = - {\pi^3 b^6 V \over 16 G_5} \ {\eta \over \beta^3}
\ee
and the soliton free energy is given by,
\be\label{fs}
F_S = - {\pi^3 b^6 V \over 16 G_5} \  {\beta_S \over \eta_S^3}, 
\ee
where, we have used the global AdS as a reference point in both cases.
Boundary topology must be the same for the black hole spacetime, the
soliton spacetime and the global AdS spacetime, in order to compare
them.  Now if we compare the free energy of black hole spacetime and
soliton spacetime then we can see that depending on the relative size
of boundary ${\bf S^1}$ circles, the bulk spacetime is dominated by
either black hole phase or the AdS soliton phase.  So, in order to
find possible phase transition between the black hole and the AdS
soliton we have to compute the difference between black hole on-shell
action (or free energy) and soliton on-shell action (or free energy).

\begin{figure}
\centering
\includegraphics[height=6cm]{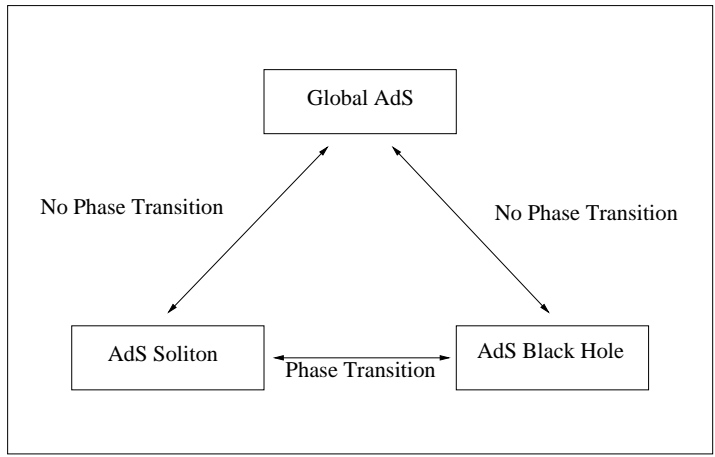}
\caption{Black Hole - Soliton Phase transition}
\label{chart}
\end{figure}

The next step is to compute the
difference between the on-shell black hole action and on-shell soliton
action with the condition that the asymptotic boundary geometry of the
black hole spacetime is same with that of soliton spacetime. This
identification implies that,
\ben\label{geoident}
\beta \sqrt{g_{\tau_B \tau_B}(\tilde R)} &=&  \beta_S \sqrt{g_{\tau_S \tau_S}
(\tilde R)},\nonumber \\
\eta_S \sqrt{g_{\theta_S \theta_S}(\tilde R)} &=& \eta
\sqrt{g_{\theta_B \theta_B}(\tilde R)},
\een
where, $\tilde R (\rightarrow \infty)$ is the position of the boundary
hypersurface. Another important thing we should mention here is that
integration over the radial coordinate $r$ ranges from $r_0$ to
$\tilde R$ for black hole spacetime and $\mu$ to $\tilde R$ for
soliton spacetime.  We will eventually take limit $\tilde R
\rightarrow \infty$ at the end of our calculations. We can now
calculate the regularized or subtracted action, which is given by,
\ben\label{subacn}
I &=& \left[ I_B(\tilde R) - I_S(\tilde R) \right]_{\tilde R \rightarrow 
\infty} 
\nonumber \\
&=& {\beta \ \eta_S \ V \over 16 \pi G_5} \lb {\mu^4 \over b^2} - {
  r_0^4 \over b^2} \rb, 
\een 
where 
\be\label{V}
V = \int \sqrt{h} dx^1 dx^2. 
\ee
Hence the free energy is,
\ben\label{freee}
F &=& \beta^{-1} I  \nonumber \\
&=& { \eta_S \ V \over 16 \pi G_5} \lb {\mu^4 \over b^2} - { r_0^4 \over b^2} 
\rb\nonumber \\
&=& { \eta_S \ V \over 16 \pi G_5} {r_0^4 \over b^2}\lb {\mu^4 \over
  r_0^4} - 1 \rb.  
\een 
We see that free energy is negative for
$r_0>\mu$ and positive for $r_0<\mu$, signaling a phase transition.
One can also express the free energy in terms of the temperature and
the horizon area (size) of the black hole. It is easy to find the
ratio, 
\be\label{size-tem-ratio} 
{{\cal A}_B \over T^2} \sim {r_0
  \over \mu}, 
\ee 
where ${\cal A}_B$ is the area of the black hole horizon.  It is
obvious from equations \ref{freee} and (\ref{size-tem-ratio}) that the
phase transition depends not only on the temperature of the black
holes but also on the size of the black holes unlike the usual HP
transition for spherical black holes where it completely depends on
the temperature.  The result can be summarized as follows. There are
three phases in the bulk spacetime corresponding to a given boundary
topology: the black hole phase, the global AdS phase and the AdS soliton
phase. First and third phase are always dominant over the second one.
But when boundary Euclidean time circle is less than the size of
spatial ${\bf S^1}$ the black hole phase dominates in bulk otherwise
the AdS soliton phase dominates, see figure \ref{chart}.

Although discussing the phase transition was the main goal of this
section, for completeness we will also write the expressions for
thermodynamic variables, energy and entropy. Using the following
thermodynamic relations,
\ben\label{nutral-mass-entropy-frmla}
E&=& {\p I \over \p \beta},\nonumber \\
S &=& \beta E - I,
\een 
we find the energy and entropy as follows,
\ben\label{nutral-mass-entropy}
E&=& {V \eta_S \over 16 \pi G_5 b^2} \lb 3 r_0^4 + \mu^4 \rb, \nonumber \\
S&=& {{\cal A}_B \over 4 G_5}.
\een
It is clear that energy is always positive with respective to the soliton.

We will conclude this section by mentioning the correspondence between
this phase transition (in the bulk) and the phase transition of dual
gauge theory on the boundary. For complete discussion the reader is
referred to the original paper \cite{sumati1}. The stable black hole
phase $r_0 > \mu$ corresponds to de-confined phase of the thermal
gauge theory on ${\bf S^1 \times S^1 \times R^2}$ and the stable
soliton phase $r_0 < \mu$ corresponds to confined phase of the thermal
gauge theory on ${\bf S^1 \times S^1 \times R^2}$. It is important to
mention here that only the antiperiodic spin structure ($Tr e^{-\beta
H}$) in both the ${\bf S^1}$ directions undergoes the large $N$ phase
transition.



\sectiono{Phase Transition of Charged Black Holes }\label{chargebh}

So far we have seen that neutral Ricci-flat AdS black holes undergo a
phase transition into a global soliton spacetime. Therefore, it would
be interesting to generalize this idea to the case of charged black
holes.

Electrically charged black holes in five dimensions have drawn a lot
of interest in the context of AdS/CFT. The electric charge of these
black holes are mapped to the global R-charge of the dual field
theory. Because of the presence of the electric charges, the
thermodynamics and the phase structure of these black holes are rather
complicated and also interesting at the same time. There have been
a lot of study of thermodynamics and phase transitions of these
charged black hole with different horizon topologies (see
\cite{cvetic1,jonson,cai} and references therein). But it seems that
the study of thermodynamics and phase transition between charged
Ricci-flat black holes and AdS soliton spacetime are yet to be
explored. In this section we will shed some light on these
issues.

We will first briefly discuss how charge black holes can arise in five
dimensions in the context of string theory.  A consistent truncation
of ${\cal N}=8$, $D=5$ gauged supergravity with $SO(6)$ Yang-Mills
gauge group, which can be obtained by $S^5$ reduction of type IIB
supergravity, gives rise to ${\cal N}=2$, $D=5$ gauge supergravity
with $U(1)^3$ gauge group.
The same theory can also be obtained by compactifying eleven
dimensional supergravity, low energy theory of M theory, on a
Calabi-Yau three folds. The bosonic part of the action of ${\cal
  N}=2$, $D=5$ gauged supergravity is given by, 
\ben\label{gauge-sugra-acn} 
I_{sugra} &=& \int d^5x \sqrt{-g} \lb {R
  \over 16 \pi G_5} +  {V(X) \over b^2} -{1 \over 2} G_{IJ}(X) \p_{\mu}
X^I \p^{\mu}X^J - {1 \over 4}
G_{IJ}(X)F^I_{\mu\nu}F^{\mu\nu \ J} \rb \nonumber \\
&+& C.S. \ \ terms , 
\een 
where, $X^I$'s are three real scalar fields, subject to the constraint
$X^1 X^2 X^3 =1$. $F^I$'s are field strengths of three Abelian gauge
fields (I,J=1,2,3). The scalar potential $V(X)$ is given by,
\be\label{sp}
V(X)= 2 \lb {1\over X^1}+{1 \over  X^2}+{1 \over X^3} \rb.
\ee
The metric on the scalar manifold $G_{IJ}$ is given by,
\be\label{scalarmet}
G_{IJ}(X) = {1 \over 2} diag \lb {1 \over (X^1)^2}, {1 \over (X^2)^2},  {1 
\over (X^3)^2} \rb.
\ee
The solution of this action is specified by the following metric,
\ben\label{sugramet}
ds^2_{sugra} = - (H_1 H_2 H_3)^{-2/3} f dt^2 + (H_1 H_2 H_3)^{1/3} {dr^2 
\over f} +  (H_1 H_2 H_3)^{1/3} {r^2 \over b^2} d\Omega_k^2, \nonumber \\
\een 
where 
\ben\label{fandH}
f = k - {M \over r^2} + {r^2 \over b^2} H_1 H_2 H_3, \ \ \ H_I = 1 + {\tilde 
q_I \over k r^2},
\een
The
three real scalar fields $X^I$'s and gauge potentials $A_{\mu}^I$'s
are of the form,
\ben\label{scalar-gauge}
X^I = H_I^{-1} (H_1 H_2 H_3)^{1/3}, \ \ \ \ A_t^I = -{\sqrt{k \tilde q_I
(\tilde q_I + M)} \over k r^2 + \tilde q_I} + \Phi^I,
\een
where $\Phi^I$'s are constants.  $k=1,0,-1$ corresponds to black holes
with spherical, Ricci-flat and hyperboloid horizon topology
respectively. We are interested in k=0 case. This is done by taking the 
limit $k \ra 0$ , $\tilde q_I \ra 0$ with $q_I = \tilde q_I/k$ fixed. Then 
the solution becomes,
\ben\label{flatsol}
ds^2_{sugra,k=0} = - (H_1 H_2 H_3)^{-2/3} f dt^2 + (H_1 H_2 H_3)^{1/3} {dr^2 
\over f} &+&  (H_1 H_2 H_3)^{1/3} {r^2 \over b^2} d\theta^2 \nonumber \\
&+& (H_1 H_2 H_3)^{1/3} {r^2 \over b^2}d\vec x^2,\nonumber \\
f=- {M \over r^2} + \tilde g^2 r^2 H_1 H_2 H_3, \ \ \ X^I = H_I^{-1} (H_1 H_2 
H_3)^{1/3}, &&A_t^I = -{\sqrt{ q_I M} \over  r^2 + q_I}+\Phi^I,
\een
where $\theta$ is an angular coordinate with period $\eta$. We will
choose $\Phi^I$'s in such a way that the gauge potential vanishes on
the horizon $r_0$,
\be\label{phi}
\Phi^I = {\sqrt{q_I M}\over r_0^2 + q_I},
\ee
where horizon radius $r_0$ is given by the solution of the following 
equation,
\be
f(r_0) =0.
\ee
It turns out that asymptotic values $\Phi_I$'s of gauge fields 
$A_I$'s behave as 
chemical potential when we consider the black hole thermodynamics in 
grand canonical ensemble.
 
At this point we would like to add a remark that $k=0$ three charge
AdS black holes of ${\cal N}=2$ gauged supergravity in $D=5$ can be
embedded in $D=10$ as a solution that is precisely the decoupling
limit of the rotating $D3$ brane \cite{jonson,tenauth}.

One special case, namely three equal charge case ($q_1=q_2=q_3=q$), of
this black hole solution is actually the same as the
\adsrn \ solution described in equations
(\ref{bhmet})-(\ref{gaugepot}).  In this case scalar fields become
constant and the scalar potential in the action reduces to a constant
value.  The action becomes the Einstein-Maxwell action in the AdS
spacetime. Though the metric looks somewhat different than usual
\adsrn \ metric, after a suitable coordinate transformation one can
write this metric in usual \sh \ coordinate. It is important to
mention here that although \adsrns \ and three equal charge R-charged back
holes show the same kind of phase transition, their local stability
behavior is somewhat different from each other. We will discuss these
issues later in this paper.

In the next subsection we discuss phase transition and thermodynamics
of Ricci-flat \adsrn, \ which is somewhat simpler than general
R-charge black hole, and will be a warm-up example. In subsection
\ref{rcharge} we will focus on general R-charge black holes.


\subsection{\adsrns} \label{rn}

\adsrns \ are the solution of field equations governed by the
Einstein-Maxwell action with negative cosmological constant. The
action is given by,
\be \label{chargebhacn}
I = - {1\over 16 \pi G_5}\integrat \left(R+{12 \over
  b^2} -F^2\right).
\ee
The solution is specified by the a metric and a gauge field. The metric is 
given by,
\be\label{bhmet}
ds^2 = -V(r) dt^2 + {dr^2 \over V(r)} + {r^2 \over b^2} d \theta^2 + {r^2 
\over b^2} (dx_1^2 + dx_2^2),
\ee
where $V(r)$ is given by,
\be\label{bhV}
V(r) = - {m \over r^2}+ {q^2 \over r^4} + {r^2 \over b^2},
\ee
and $m$ and $q$ are two constants of integration (parameters). Later we will 
relate them to the ADM mass and physical charge of the black hole 
respectively. 
The gauge field is given by,
\be\label{gaugepot}
A(r) = \lb -{1 \over c} {q \over r^2} + \Phi \rb dt,
\ee
where $c= 2/\sqrt{3}$ and $\Phi$ is constant. We will choose $\Phi$ in such 
a way that $A(r_0) = 0$ and this gives 
\be
\Phi={1 \over c} {q \over r_0^2}.
\ee
where $r_0$ is the position of horizon and is given by the solution of the 
following equation, $$V(r_0) =0.$$
The coordinate $\theta$ is periodic and has a period $\eta$. This five 
dimensional spacetime (X) has a conformal boundary $M=R \times S^1 \times 
R^2$.

In order to consider thermodynamics of this black hole in the classical limit 
we will work with the Euclidean theory which is obtained by a Wick rotation $t 
\rightarrow i \tau$. The Euclidean metric is given by,
\be\label{ebhmet}
ds^2 = V(r) d\tau^2 + {dr^2 \over V(r)} + {r^2 \over b^2} d \theta^2 + {r^2 
\over b^2} (dx_1^2 + dx_2^2)
\ee
and the gauge field becomes,
\be\label{egaugepot}
A_{\tau} = i \lb -{q \over c r^2} + {q \over c r_0^2} \rb
\ee
The Hawking temperature of the black hole is given by,
\be\label{bhtemprn}
T = {2 r_0^6 -q^2 b^2 \over 2 \pi b^2 r_0^5}
\ee
We will now calculate the on-shell black hole action, (\ref{chargebhacn}),
which is given by,
\be \label{onshellbhacn}
I_{B} = {\beta \ \eta \ V \over 16 \pi G_5 b^3} \lb {2(\tilde R^4 - r_0^4) 
\over b^2} - {2 q^2 \over r_0^2} \rb,
\ee
where $\tilde R$ is the cutoff in the radial direction and $V=\int
dx_1 dx_2$ (\ref{V}). As $\tilde R \rightarrow \infty$, the on-shell
action equation (\ref{onshellbhacn}) diverges. To study the phase
transition between the black hole and AdS soliton we will subtract the
contribution of the AdS-soliton from black hole action and it will
also regularize the onshell action.

As we mentioned earlier given the boundary topology to be $R 
\times S^1 \times R^2$
 (Lorentzian), the AdS-Soliton is conjectured to be the minimum energy
 solution of the action (\ref{chargebhacn}). 
The solution is given by a
constant gauge field and a metric (\ref{adssol1}),
\be 
ds_S^2= -dt_S^2 + {dr^2 \over V_S(r)} + V_S(r) d\theta_S^2 + {r^2 \over b^2} 
(dx_1^2 + dx_2^2)\nonumber
\ee
where
\be\label{solitonV}
V_S(r) = {r^2 \over b^2} \lb 1 - {\mu^4 \over r^4} \rb.
\ee
The coordinate $\theta_S$ has period,
\be\label{solitonperiod}
\eta_S = 
{\pi b^2 \over \mu}.
\ee
The Euclidean soliton metric is given by (\ref{euclisol}),
\be\label{euclisolmet}
ds_S^2 = {r^2 \over b^2}d\tau_S^2 + {dr^2 \over V_S(r)} + 
V_S(r) d\theta_S^2 + {r^2 \over 
b^2} (dx_1^2 + dx_2^2),
\ee
where $\tau_S$ can have an arbitrary period $\beta_S$. Now we can calculate 
the on shell AdS-Soliton action which is given by,
\be\label{onshellsolacn}
I_S={\beta_S \ \eta_S \ V \over 16 \pi G_5 b^3} \lb {2(\tilde R^4 - \mu^4) 
\over b^2} \rb.
\ee
Before subtracting the AdS-Soliton contribution from the black hole
contribution we should remember that in order to match the boundary
geometry of black hole spacetime to that of the soliton spacetime we
have the relations (\ref{geoident}) \footnote{In addition to these
relations we also have to identify the chemical potential of black
hole spacetime to that of soliton spacetime. Since AdS soliton is a
solution of the Einstein equations with constant gauge potential, we can
assign any chemical potential for this spacetime.}, which in the
present case reduces to 
\ben\label{chargetempreletion} \beta
\sqrt{V(\tilde R)} ={\tilde R \over b} \beta_S, \nonumber \\ \beta_S =
\beta \lb 1 - {m \ b^2 \over 2 \tilde R^4} \rb \een and
\ben\label{chargeetarelation} \eta_S \sqrt{V_S(\tilde R)} = {\tilde R
\over b}\eta ,\nonumber \\ \eta= \eta_S \lb 1 - {\mu^4 \over 2 \tilde
R^4} \rb.  \een 
Using these relations we can now calculate the
regularized or subtracted action, 
\ben \label{chargesubacn} I &=&
\left[ I(\tilde R) - I(\tilde R) \right]_{\tilde R \rightarrow \infty}
\nonumber \\
 &=& {\beta \ \eta_S \ V \over 16 \pi G_5} \lb {\mu^4
\over b^2} - {2 q^2 \over r_0^2} - {2 r_0^4 \over b^2} + m \rb
\nonumber \\ 
&=& {\beta \ \eta_S \ V \over 16 \pi G_5} \lb {\mu^4
\over b^2} - {q^2 \over r_0^2} - { r_0^4 \over b^2} \rb.  
\een 
And hence the Gibbs free energy is given by, 
\ben\label{chargefreeenergy}
G &=& T I \nonumber \\ 
&=& {\eta_S \ V \over 16 \pi G_5} \lb {\mu^4
\over b^2} - {q^2 \over r_0^2} - { r_0^4 \over b^2} \rb.  \een


\subsubsection{Thermodynamics} \label{thermorn}

In this subsection we will discuss the thermodynamics of \adsrns.
There are two kinds of ensembles one can use to describe the
thermodynamics of a charged black hole system. $(i)$ Canonical
ensemble - where temperature of the system is fixed but energy can
flow between the system and the heat bath. $(ii)$ Grand canonical
ensemble - where temperature and the electric potential are fixed and
energy and charge can flow.

In this paper we will study thermodynamics considering the system in a
grand canonical ensemble, since it is interesting in the context of
AdS/CFT.

A state in grand-canonical ensemble is characterized by (inverse)
temperature $\beta$ and electric potential $\Phi$. In the \gc the
Gibbs potential is given by,
\be\label{gibbs}
G = {I \over \beta} = E - T S- \Phi Q,
\ee
where $E$ is the energy, $S$ is the entropy and $Q$ is physical charge
of the system. Using the above relation we may compute the variables
of the system as follows,
\ben \label{EQS}
E &=& \lb {\p I \over \p \beta} \rb_{\Phi} - {\Phi \over \beta} \lb {\p I 
\over \p \Phi} \rb_{\beta}, \nonumber \\
S &=& \beta \lb {\p I \over \p \beta} \rb_{\Phi} -I \nonumber, \\
Q &=& -{1 \over \beta}\lb {\p I \over \p \Phi} \rb_{\beta}.
\een  
Together, they satisfy the first law,
\be\label{firstlaw}
dE = T dS + d(\Phi Q).
\ee
Using the Euclidean action  (equation \ref{chargesubacn}) and the 
above relations we 
compute the thermodynamic variables as follows,
\ben\label{Evalue}
E={V \eta_S \over 16 \pi G_5}\lb 3 m + {\mu^4 \over b^2}\rb,
\een
\ben\label{Svalue}
S={1 \over 4 G_5} {V \eta_S r_0^3 \over b^3} = {A \over 4 G_5},
\een
\ben\label{Qvalue}
Q_{phy}= {\sqrt{3} \over 4 \pi G_5} {V \eta_S \over b^3} q.
\een
Equation (\ref{Evalue}) shows that the energy is always positive with 
respect to the soliton energy. 

\begin{figure}
\centering
\includegraphics[height=6cm]{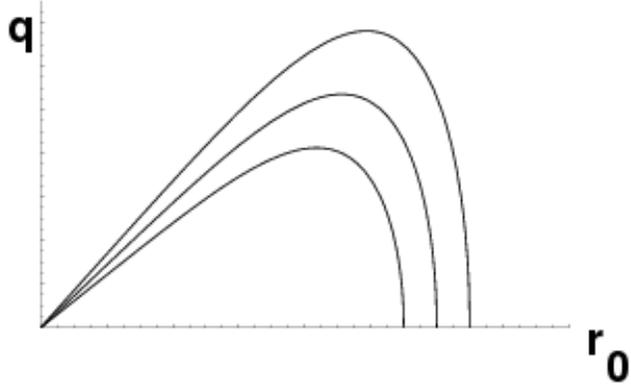}
\caption{Phase transition curves (equation \ref{PTcond:qr0}) in $q-r_0$ 
plane for \adsrn.\ 
Three different curves for three different values of $\mu$. As we increase
the value 
of $\mu$ corresponding phase transition curves go away from the 
$r_0$ axis.} 
\end{figure}


\subsubsection{Phase Transition} \label{ptrn}
Clearly the Gibbs free energy (\ref{chargefreeenergy}) carries the 
signature of phase transition.
When $G< 0$ the black hole phase dominates. On the other hand,
when $G>0$, the black hole phase is unstable and decays to the
soliton.  Once we fix the size of the
boundary spatial circle, {\it i.e.}, fixed $\eta$ (and so fixed $\eta_S$ or
$\mu$), the phase transition curve depends on the charge '$q$' and the
size of the black hole '$r_0$'. Conditions are given by the following
relations, \ben \label{PTcond:qr0}
r_0^6 - \mu^4 r_0^2 + q^2 b^2 > 0 \ \ \ \ \ \ \  \ black \ \ hole \ \ phase 
\ \ dominates, \nonumber \\
r_0^6 - \mu^4 r_0^2 + q^2 b^2 < 0 \ \ \ \ \ \ \ \ AdS\ soliton \ phase
\ dominates.  \een

Since a system in \gc is specified by its temperature and chemical potential,
it would be interesting to plot the phase transition diagram in $T-\Phi$ 
plane. For a given \cmp , the temperature can be written as,
\be
T = {3 r_0^2 -2 b^2 \Phi^2 \over 3 \pi b^2 r_0}.
\ee

\begin{figure}
\centering
\includegraphics[height=6cm]{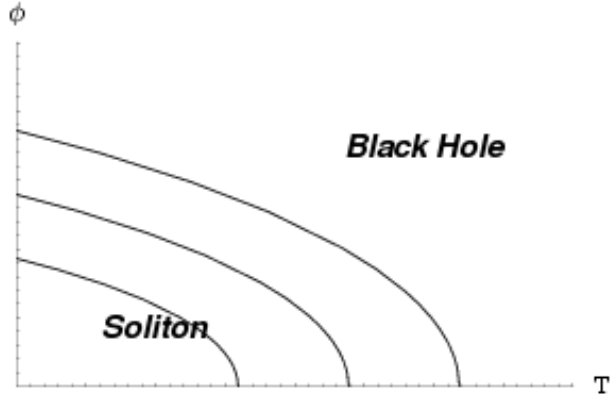}
\caption{$T_{tran}$ Vs $\Phi$ Plot for \adsrn.\ Different curves 
correspond to different 
$\mu$. For smaller value of $\mu$ the corresponding phase transition
curves are closer to origin. When $T>T_{tran}$ BH spacetime has 
dominant contribution to Euclidean path integral}
\label{figTphi}
\end{figure}

Using the relation \ref{PTcond:qr0}, we can write the relation between 
transition temperature $T_{tran}$, \cmp and $\mu$,
\ben\label{Tphimu}
T_{tran} &=& \sqrt{{-4b^2 \Phi^2 + \sqrt{16 b^4 \Phi^4 + 36 \mu^4} \over 6}}
{1 \over \pi b^2} \nonumber \\
&& - {2 \Phi^2 \over 3 \pi }\lb {-4b^2 \Phi^2 + \sqrt{16 b^4 \Phi^4 + 36 
\mu^4} \over 6} \rb^{-1/2}.
\een
Fig. \ref{figTphi} depicts how transition temperature changes with \cmp for 
different $\mu$.

We will end this subsection briefly mentioning the effect of \gb \ to
the action \ref{chargebhacn}. The solution is give in
\cite{torii}. Since \gb \ does not change the thermodynamics (energy,
entropy and physical charge) for Ricc-flat black holes we are not
presenting the details of our result here. Phase transition
condition also does not receive any correction in presence of the \gb
\ in the action. One can check that the functional dependence of free
energy (which is proportional to difference between on shell black hole
action and on shell soliton action) on charge $q$ horizon radius $r_0$
and radius of boundary spatial circle is same as equation
\ref{chargefreeenergy}, only the over all coefficient changes.


\subsection{R-charge Black Holes} \label{rcharge}

In this subsection we will focus on phase transition of $D=5$ R-charge
black holes.
First we will give a general expression of free energy for any
arbitrary three charge case.  In order to study the phase transition
and stability we will focus on three different limits, $(i)$ one
charge case, {\it i.e.}, $q_1=q,q_2=q_3=0$, $(ii)$ two equal charge case,
$q_1=q_2=q,q_3=0$ and $(iii)$ three equal charge case,
$q_1=q_2=q_3=q$.

In order to calculate free energy for these black
holes we will follow same steps as in the last subsection.  But here 
we will encounter a
subtlety. In the previous case both the black hole and the soliton
metric were written in \sh \ coordinate and hence we had calculated
both the on-shell action and the background action keeping in mind
relations (\ref{chargetempreletion}) and (\ref{chargeetarelation}).
But in this case the black hole metric (\ref{sugramet}) is written in the
``isotropic coordinates'' but the soliton metric (\ref{adssol1}) is
written in the usual \sh coordinates. So either we have to write the black
hole metric in the \sh \ coordinate by suitably changing the coordinates
and redefining the parameters or we can write the soliton metric in the
isotropic coordinate by redefining the coordinate as follows,
\be\label{coordchange}
r^2 \ra r^2 (H_1 H_2 H_3)^{1 \over 3}.
\ee
We shall follow the second approach. After this coordinate change 
the soliton metric becomes,
\ben\label{solmetiso}
ds_S^2 &=& -{r^2 \over b^2} (H_1 H_2 H_3)^{1/3} dt_S^2 + {b^2 \over r^2} 
\lb 1 - {{q1 \over H_1}+ {q2 \over H_2}+{q3 \over H_3} \over 3 r^2} \rb^2 
\lb 1 - {\mu^4 \over r^4 (H_1 H_2 H_3)^{2/3}} \rb^{-1} dr^2 \nonumber \\
&+& {r^2 \over b^2} (H_1 H_2 H_3)^{1/3} \lb 1 - {\mu^4 \over r^4 (H_1 H_2 
H_3)^{2/3}} \rb d\Theta_S^2 +  {r^2 \over b^2} (H_1 H_2 H_3)^{1/3} d\vec x^2.
\een

The periodicity $\eta_S$ of the compact direction $\theta_S$ remains
same as (\ref{soltemp}). We will now compute the on-shell black hole
action (\ref{gauge-sugra-acn}) and subtract contribution of the global
soliton action from this to get finite or renormalized on-shell
action. In this case, unlike the \rn \ spacetime, the Gibbons-Hawking
boundary (GH) terms also give a finite contribution to the on-shell
action.

To study thermodynamics and phase transition we have to go to the
Euclidean space by Wick rotation $t \ra i \tau$.
The inverse Hawking temperature of the black hole is given by,
\be\label{temp}
\beta = {2 \pi b^2 r_0^2 \sqrt{(r_0^2 + q_1)(r_0^2 + q_2)(r_0^2 + q_3)} 
\over 2 r_0^6 + r_0^4(q_1 + q_2 + q_3) -q_1 q_2 q_3}.
\ee

Detailed derivation of the on-shell action has been given in 
appendix \ref{appA}. 
Here we will only write the result (from here we will work
in $b=1$ unit for simplicity).
\be \label{rchargeonshellacn}
I = - {\beta V \eta_S \over 16 \pi G_5} \lb M + {2 \over 3} ((q_1^2 + q_2^2 
+q_3^2) - (q_1 q_2 + q_2 q_3 + q_3 q_1)) -\mu^4 \rb.
\ee
Using this regularized action if we calculate mass of these black
holes, an unexpected nonlinear term involving charges appears
\cite{buchel}. There is nothing inherently wrong in the nonlinear
appearance of charge in the expression of mass but it can contradict
the expected BPS inequality between the charge and the mass. Liu and Sabra
\cite{sabra+liu} pointed out that inclusion of a finite counterterm
can resolve that problem. The counterterm they proposed is,
\be\label{finitecounterm}
I_{C}= {1 \over 8\pi G_5} \int d^4x \sqrt{-h} \vec \phi^2,
\ee
where $\vec \phi$ is related to three scalar fields in the following way,
$$
X_I = e^{-{1 \over 2}\vec a_I. \vec \phi},
$$
with the condition, $\vec a_I . \vec a_J = 4 \delta_{IJ} - 4/3$
\cite{tenauth}. It is straightforward to check that using this field
redefinition, the counterterm becomes,
\be
I_C = {\beta V \eta_S \over 16 \pi G_5} \lb {2 \over 3} ((q_1^2 + q_2^2 
+q_3^2) - (q_1 q_2 + q_2 q_3 + q_3 q_1)) \rb.
\ee 
After adding this counterterm we can get rid of that nonlinear charge 
term in the action. So the final action is given by,
\ben\label{rchargeonshellacnfinal}
I= - {\beta V \eta_S \over 16 \pi G_5} \lb {M } - {\mu^4 } \rb 
\een  
and the Gibbs free energy is given by,
\ben\label{free}
G&=&  - {V \eta_S \over 16 \pi G_5} \lb M - \mu^4 \rb \nonumber \\
&=& - {V \eta_S \over 16 \pi G_5} \lb r_0^4 H_1(r_0) H_2(r_0) H_3 \lb r_0 
\rb -\mu^4 \rb.
\een

\subsubsection{Thermodynamics} \label{rchargethermo}

We will use the set of thermodynamic relations given by equation
(\ref{EQS}) and Gibbs potential (\ref{gibbs}) to compute energy,
entropy and physical charge of R-charge black holes.
\be\label{Evaluer}
E = {V \eta \over 16 \pi G_5} \left [ 3 M + \mu^4 \right ],
\ee
\be\label{Svaluer}
S= {V \eta \over 4 G_5} \sqrt{(r_0^2 +q_1)(r_0^2 + q_2)(r_0^2 + q_3)} = 
{Area \over 4 G_5},
\ee
\be\label{Qvaluer}
Q^I_{phys} = {V \eta \over 8 \pi G_5}  \sqrt{q_I(r_0^2 + q_1)(r_0^2 + q_2)
(r_0^2 + q_3)/r_0^2}.
\ee

\begin{figure}
\centering
\includegraphics[height=6cm]{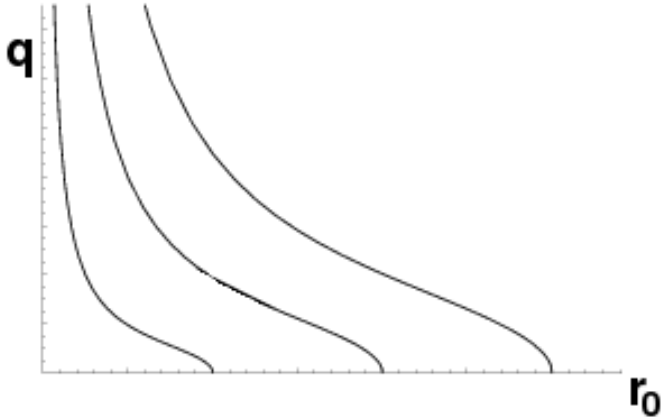}
\caption{Phase transition curves (equation \ref{pt1}) for different $\mu$ in 
$q-r_0$ plane for single charge black holes. As value of $\mu$ increases
transition curves recede from the origin.}
\label{figpt1}
\end{figure}

\subsubsection{Phase Transition} \label{rchargept}

Signature of phase transition is obvious from the expression of Gibbs free
energies (equation \ref{rchargeonshellacnfinal}) and the phase transition 
depends on the size of compact
dimension. The phase transition conditions are given by,
\ben\label{ptcondition}
r_0^4 H_1(r_0)H_2(r_0) H_3(r_0) &>& \mu^4  \ \ \ \ \ Black\ \ Hole\ \ 
Phase\ \ dominates, \nonumber \\
r_0^4 H_1(r_0)H_2(r_0) H_3(r_0) &<& \mu^4  \ \ \ \ \ Soliton\ \ Phase\ 
\ dominates.
\een

We will now discuss the phase diagram for this black hole and concentrate 
on three special cases.\\
\bc
\underline{{\bf A.} ${\bf q_1=q}$, ${\bf q_2=q_3=0}$}
\ec
In this case the phase transition conditions are given by,
\ben\label{pt1} 
r_0^4 + q r_0^2 - \mu^4 &>& 0 \ \ \ \ Black\ \ Hole\ \ Phase\ \ dominates, 
\nonumber \\
r_0^4 + q r_0^2 - \mu^4 &<& 0 \ \ \ \ AdS \ \ Soliton \ \ Phase \ \ dominates. 
\een
Figure \ref{figpt1} shows the phase transition curves for different values 
of $\mu$ in $q-r_0$ plane.

\begin{figure}
\centering
\includegraphics[height=6cm]{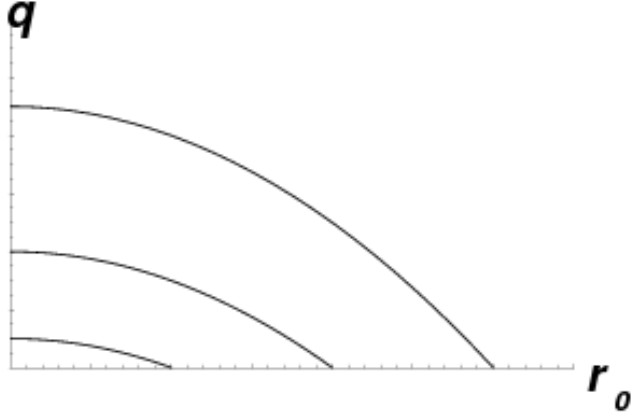}
\caption{Phase transition curves (equation \ref{pt2}) for different $\mu$ in 
$q-r_0$ plane for two equal charge black holes. }
\label{figpt2}
\end{figure}

\begin{figure}
\centering
\includegraphics[height=6cm]{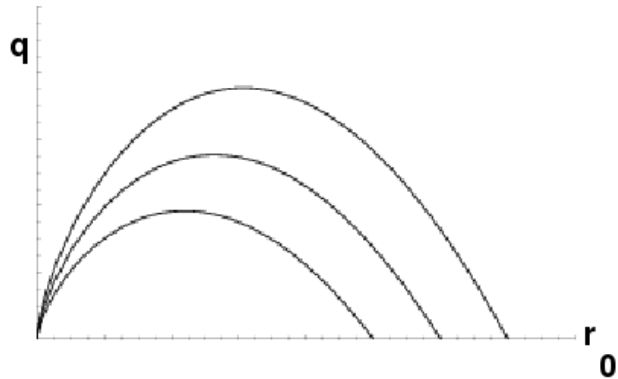}
\caption{Phase transition curves (equation \ref{pt3}) for different $\mu$ in 
$q-r_0$ plane for three equal charge charge black holes.}
\label{figpt3}
\end{figure}

\bc
\underline{{\bf B.} ${\bf q_1=q}$, ${\bf q_2=q}$ and ${\bf q_3=0}$}
\ec
For two equal charges phase transition condition is given by equation, 
\ben \label{pt2}
q &>& \mu^2 - r_0^2 \ \ \ \ \  Black\ \ Hole\ \ Phase\ \ dominates,
 \nonumber \\
q &<& \mu^2 - r_0^2  \ \ \ \ \ AdS \ \ Soliton \ \ Phase \ \ dominates. 
\een
Phase transition curves have been plotted in figure \ref{figpt2}.\\
\bc
\underline{{\bf C.} ${\bf q_1=q}$, ${\bf q_2=q}$ and ${\bf q_3=q}$}
\ec
And finally for three equal charge black holes phase transition condition 
is given by,
\ben\label{pt3}
q&>&\mu^{4/3} r_0^{2/3} - r_0^{2}\ \ \ \ \  Black\ \ Hole\ \ Phase\ \ 
dominates, \nonumber \\
q&<&\mu^{4/3} r_0^{2/3} - r_0^{2} \ \ \ \ \ AdS \ \ Soliton \ \ Phase \ \ 
dominates. 
\een
Phase transition curves have been plotted in figure \ref{figpt3}.



\sectiono{Local Stability of Charged Ricci-flat Black Holes} \label{stab}

Finally we discuss local thermodynamic stability of charged
black holes we were considering in the previous section. Local
thermodynamic stability of a system implies that the entropy $S$ which
is a function of other extensive thermodynamic variables $x_i$'s of
the system is subadditive in a sufficiently small neighborhood of a
given point in the phase space of $x_i$'s. The criterion of
subadditivity is,
\be\label{suadd}
S(\lambda x_i + (1-\lambda) x_i ) \geq \lambda S(x_i) + (1-\lambda) S(x_i), 
\ \ \ 0 \leq \lambda \leq 1.
\ee
If the inequality goes the other way, then the system can gain entropy by
dividing into two parts, one with a fraction $\lambda$ of the energy,
charge, etc and the other with a fraction $1-\lambda$. But no such
process is allowed by second law of thermodynamics, and hence
the system is thermodynamically stable \cite{cvetic1,cvetic2}.
When $S$ is a smooth function of $x_i$'s then sub-additivity is
equivalent to the Hessian matrix $\left [ {\p^2 S \over \p x_i \p x_j}
\right ] $ being negative definite.

Now one has to decide among all the extensive quantities $x_i$'s which
are thermodynamic variables of the system, {\it i.e.}, they can vary
in an experiment and which are the fixed parameters of the system.
For example, in the canonical ensemble, mass (energy) $M$ is the only
thermodynamic variable, {\it i.e.}, the system can interchange energy
with the heat bath but temperature and charges are the fixed
quantities. A canonical system is specified by its temperature and
charges. For grand canonical ensemble mass (energy) $M$ and charges
$Q_I$'s are thermodynamic variables, {\it i.e.}, the system can
interchange energy and charge with the heat bath and temperature and
chemical potential are constant parameters. In this case, the phase
space is specified by mass and charge.

So in the grand canonical ensemble the lines of instability in the phase
space are determined by finding the zeros of the determinant of the
Hessian sub-matrix $\left [ {\p^2 S \over \p x_i \p x_j} \right ] $,
where, $x_i$'s are mass and charges. For the \rn \ black holes, the Hessian
is a $2 \times 2$ matrix and for R-charge black holes this is a $4
\times 4$ matrix. It has been argued in \cite{cvetic2} that the zeros
of the determinant of the Hessian of $S$ with respect to $M$ and
$Q_I$'s coincide with the zeros of the determinant of the Hessian of
the Gibbs (Euclidean) action,
\be\label{gacn}
I_G = \beta \lb M - \sum \limits_{I=1}^3 \Phi_I Q_I \rb -S,
\ee 
with respective to $r_0$ and $q_I$'s keeping $\beta$ and $\Phi_I$'s
fixed. Note that $q_I$'s are the charge parameters entering into the
black hole solutions where $Q_I$'s are the physical charges. Though
this criteria can figure out the instability line in the phase
diagram but it is unable to tell which sides of the phase transition
lines correspond to local stability. One can figure out the stability
region by knowing the fact that zero chemical potential and high
temperature must correspond to a stable black hole solution.

Using the procedure stated in the last paragraph we will find out the
region of stability for the black holes we have discussed. We
will first consider stability of the \rn \ black holes and then focus on 
three special cases of R-charged black holes, namely one charge, two equal
charged and three equal charged black holes, for simplicity. The
general case can also be done using the expressions for mass, charge
and entropy, given in equations (\ref{Evaluer}, \ref{Svaluer},
\ref{Qvaluer}).

\begin{figure}
\centering
{\includegraphics[height=6cm]{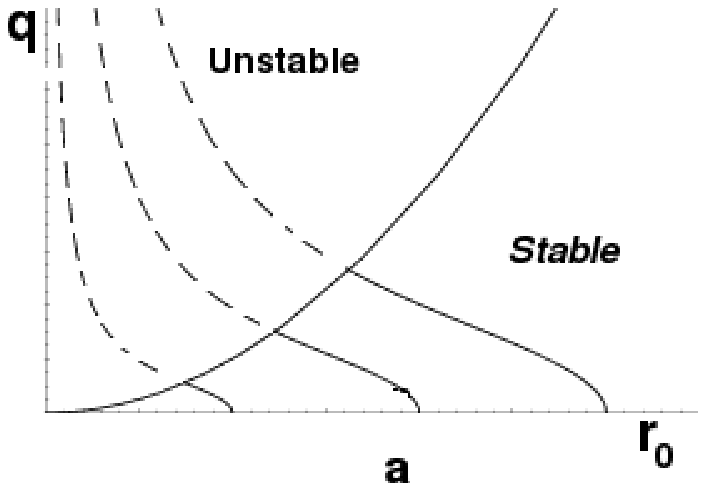}
\includegraphics[height=5cm]{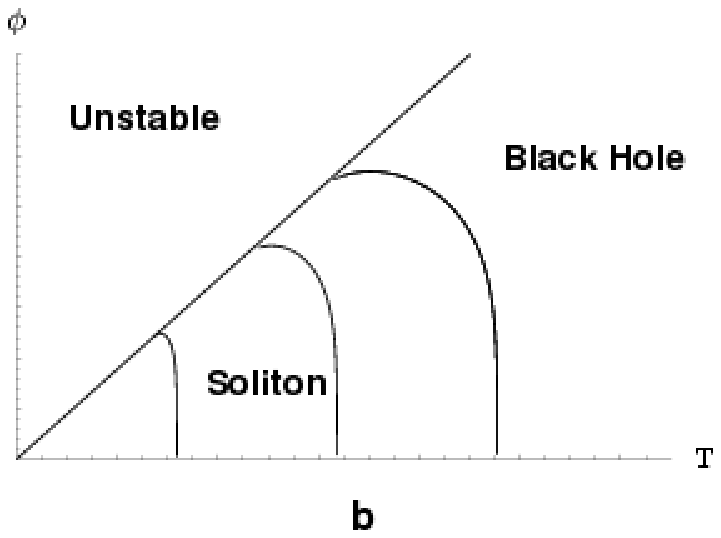}}
\caption{Fig. {\bf (a):} Phase transition curves for 
different $\mu$ and instability line in 
$r_0-q$ plane for single charge charge black holes. $q=2r_0^2$ is the 
instability line. Phase transition 
lines starts for $q=0$ and $r_0= \mu$. As we move along the phase transition 
line $r_0$ decreases and $q$ increases and when $q$ becomes equal to $2 r_0^2$
phase transition lines touch the instability lines. After that $q>2r_0^2$,
and phase transition curves enter into the unstable region (dashed lines).
Fig.{\bf (b):} Same curves have been plotted in $T-\Phi$ plane.}
\label{1ch_qr}
\end{figure}

\bc
\underline{{\bf A. \rn \ Black Holes}}
\ec
\adsrn \ is a single charge black hole. The Hessian is a $2 \times 2$
matrix. Determinant of this matrix vanishes at $r_0=0$. So in $r_0-q$
plane the line $r_0=0$ is the line of instability, and the black 
hole is stable at any $r_0>0$ and $q>0$ point .

For \rn \ black holes $q$ must be less than $2r_0^2$, otherwise the black 
hole will have negative temperature. $q=2 r_0^2$ corresponds to
$T =0$ line in $T-\Phi$ plane. So the black 
hole is stable 
at any non-zero temperature and chemical potential. No instability 
line is found in the $T-\Phi$ plane. 
\newpage
\bc
\underline{{\bf B.} ${\bf q_1=q}$, \ ${\bf q_2=q_3=0}$}
\ec
In this case also the Hessian is a $2\times 2$ matrix. 
Zeros of the determinant are given by the following condition,
\be\label{dhess1}
q=2r_0^2.
\ee
The region of stability is determined by the condition, $q<2r_0^2$
where $q=2r_0^2$ is the line of instability. We have plotted the
instability lines and phase transition line in $q-r_0$ plane as well
as in $T-\Phi$ plane in figure \ref{1ch_qr}. In the $T-\Phi$ plane the
instability line is a straight line and the region,
\be\label{tphi1} 
{\Phi \over T} < {\pi \over \sqrt 2}
\ee
is stable \cite{gubser}.

\begin{figure}
\centering
{
\includegraphics[height=5cm]{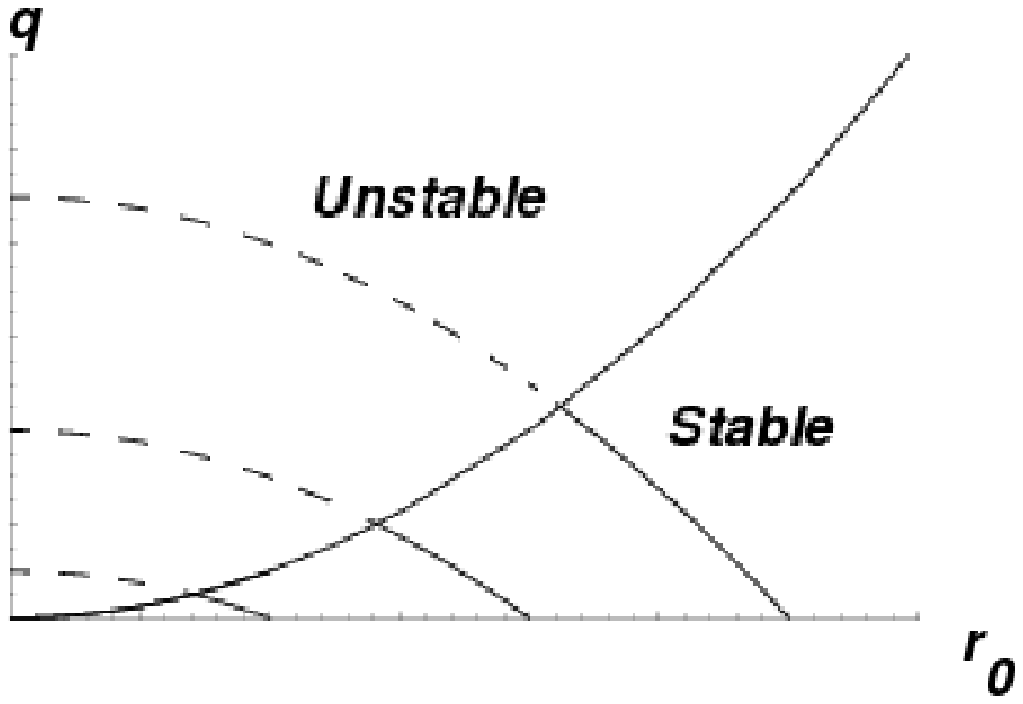} 
\includegraphics[height=5cm]{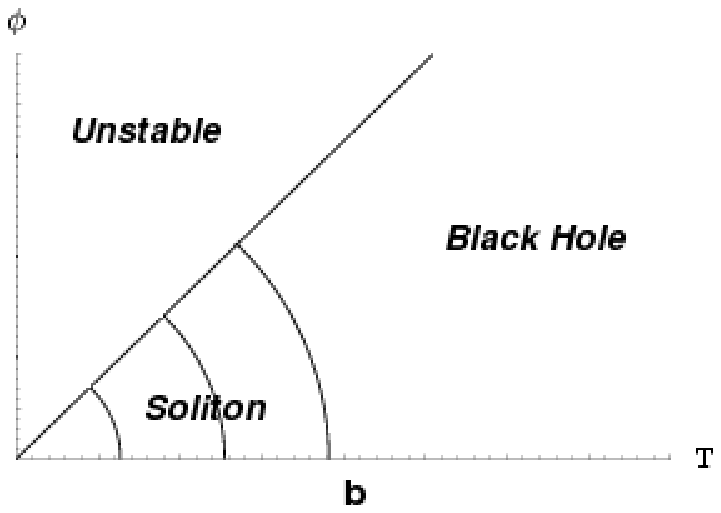}
}
\caption{Fig.{\bf (a):}Phase transition curves for 
different $\mu$ and instability line in 
$r_0-q$ plane for two equal charge charge black holes. $q= r_0^2$ is the 
instability line. Fig. {\bf (b):} 
Phase transition curves and
instability line $T -\Phi$ plane.}
\label{2ch_qr}
\end{figure}

\begin{figure}
\centering
{
\includegraphics[height=6cm]{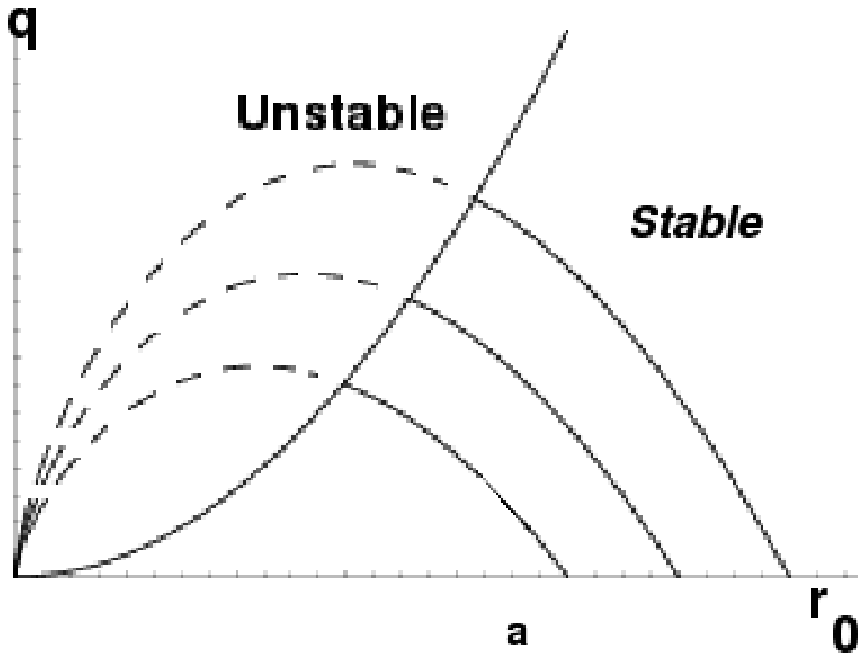}
\includegraphics[height=6cm]{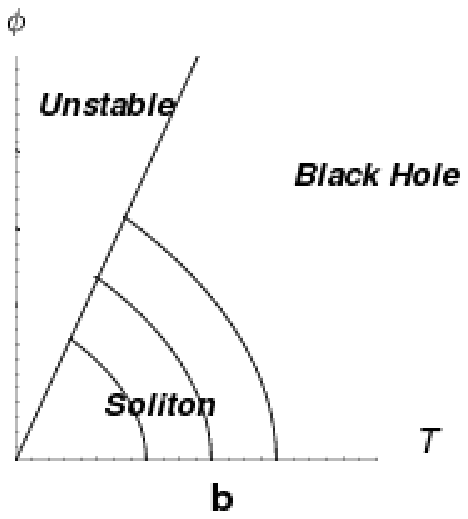}
}
\caption{Phase transition and stability curves for three equal charge black
 holes. Fig.{\bf (a):} Phase transition curves for
different $\mu$ in 
$r_0-q$ plane. Instability lines is $q=r_0^2$. Fig. {\bf (b):} 
Phase transition curves and
instability line $T -\Phi$ plane.  }
\label{3ch_qr}
\end{figure}

\bc
\underline{{\bf C.} ${\bf q_1=q}$, \ ${\bf q_2=q}$, \ ${\bf q_3=0}$}
\ec
To find out the stability of two equal charge black holes, we have to
keep it in mind that, first we will find out the Hessian matrix with
two independent non zero charges and then set them equal. So in this
case the Hessian is a $3 \times 3$ matrix. Zeros of its determinant are
given by,
\be\label{dhess2}
q=r_0^2.
\ee
The region of stability is 
\be
q<r_0^2
\ee
 in $r_0-q$ plane and 
\be
{\Phi \over T} < \pi
\ee
in $T-\Phi$ plane. See figure \ref{2ch_qr}. 

\bc
\underline{{\bf D.} ${\bf q_1=q}$, \ ${\bf q_2=0}$, \ ${\bf q_3=q}$}
\ec
Similarly for three equal charge case also we will first compute the
Hessian for three independent charges, so the Hessian is a $4 \times 4$
matrix. Then we set three charges to be equal. The determinant vanishes for,
\be\label{dhess3}
q=r_0^2.
\ee
Hence $q = r_0^2$ determines the instability line and  
\be
q<r_0^2
\ee
is the region of stability in $r_0-q$ plane. Similar plot exists in
the $T-\Phi$ plane, where stability region is determined by [figure
\ref{3ch_qr}],
\be\label{tphi3}
{\Phi \over T} < 2 \pi.
\ee


\sectiono{{\bf Discussion}}\label{dis}

We have investigated the nature of phase transition curves of charged
Euclidean black holes in five dimensions whose asymptotic boundary
topology is ${\bf S^1 \times S^1 \times R^2}$. Bulk spacetime with
this boundary topology does not show any signature of phase transition
between black hole and a global AdS spacetime.  The Gibbs free energy
remains always negative for any positive value of $r_0$ and $q$, and
hence black hole phase is always dominant in the Euclidean path integral.
But we have shown that instead of global AdS if we compare the black
hole with the AdS soliton then the black hole could undergo a phase
transition.  Depending on the size of the boundary ${\bf S^1}$ circle the
Gibbs potential flips sign as we vary $r_0$ and $q$ or temperature
($T$) and chemical potential ($\Phi$).

First we have considered the simplest
five dimensional charged black holes which are \rn \ black holes and
studied their thermodynamics and phase transition. We then focused on
various five dimensional R-charged black holes which arise as a solution of
${\cal N}=2$ gauged supergravity in five dimensions. We have shown in all 
four cases that the phase transition lines in the phase diagrams 
typically depend on the size of the boundary $S^1$ circle. As a 
consistency check, if we set $\mu$, which is proportional to 
inverse radius of boundary circle, to zero then the radius of ${\bf S^1}$ 
circle becomes infinity, the boundary space becomes ${\bf R^3}$ and 
all the phase transition lines disappear from the phase diagram, as expected.

We have also discussed the stability of these Ricci-flat charged black
holes and shown that the R-charged black holes are locally stable in
some region of the phase space. The black hole instability line rises
linearly with temperature in $T-\Phi$ plane and the slope is ${\pi
\over \sqrt{2}}$, $\pi$ and $2 \pi$ for single charge black holes, two
equal charge black holes and three equal charge black holes
respectively. The instability lines do not depend on the size of the
boundary ${\bf S^1}$ circle. It depends on topology of boundary
spacetime, {\it i.e.}, whether the the boundary spacetime is flat
$(k=0)$, spherical $(k=1)$ or hyperboloid $(k=-1)$. For spherical
black holes instability lines are found in \cite{cvetic1}.  In
\cite{cvetic1} instability lines were also found for $k=0$ black holes
(single charge case).  Our results agree with theirs.

These five dimensional Euclidean R-charged black holes with topology
${\bf {\cal B}^2 \times S^1 \times R^2}$ are dual to the Euclidean weakly
coupled field theory on ${\bf S^1 \times S^1 \times R^2}$ with three
chemical potentials turned on \footnote{Gauge theory on ${\bf R^3}$ or
${\bf S^3}$ with ${\bf U(1)}$ R-charges has been discussed in
\cite{gubser,yy}.}. The linear behavior of instability lines in phase
diagram are also expected in dual gauge theory side.  When one turns
on some non-zero chemical potential in the gauge theory, this chemical
potential acts like a negative mass squared term for the scalars
\cite{yy}. So if we consider the theory at zero temperature and zero
coupling then it does not have any stable ground state, as the
potential for the scalar fields is unbounded from bellow. But at some
finite temperature the scalars gain a thermal mass at one loop level
which is proportional to $\sqrt{\lambda} T$, where $\lambda$ is 't
Hooft coupling. And hence, as long as maximal chemical potential is
less than $ \sqrt{\lambda} T$ the theory has stable ground state. So
in the gauge theory side also we can see linear behavior of
instability line.

It would be interesting to find out confinement\ -\ deconfinement
phase transition on the weak coupling side by computing the partition
function of ${\cal N}=4$ $SYM$ theory on ${\bf S^1 \times R^2}$ with
three nonzero chemical potential \footnote{see \cite{aharony} for
  related discussion.}.  Also one has to keep in mind that the fermions
are antiperiodic along the spatial ${\bf S^1}$. It has been argued in
\cite{page} from the point of view of AdS/CFT that in the limit of
large $N$ a conformally invariant gauge theory on a flat torus (with
anti-periodic boundary conditions for the fermions in all the compact
directions) undergoes a phase transition when two shortest
periodicities are interchanged. Therefore, it seems to be interesting to
write a gauge invariant partition function for the gauge theory on
${\bf S^1 \times R^2}$ in presence of (three) chemical potential and
understand how size of the ${\bf S^1}$ circle governs
confinement-deconfinement phase transition in the weak coupling side.
\\
\bc
{\bf \underline{ \ \ \ \ \ \ \ \ \ \hskip 5cm }}
\ec
\vskip 1.3cm

\noindent
{\bf Acknowledgement}

We are thankful to R.Gopakumar, D. Jatkar, A. Sen 
for helpful discussions and their valuable comments on our first manuscript.
We are also thankful to D. Astefanesei for helpful discussions. 

\newpage
\noindent
{\Large \bf Appendix}
\appendix
\sectiono{Calculation of Free energy for R-charged Black Holes} \label{appA}

In order to calculate the on-shell action we will write the action 
\ref{gauge-sugra-acn} using equations of motion as,
\be\label{onshellacn} 
I_B^{bulk} = -{1 \over 8 \pi G_5} \integrat R^{\theta}_{\theta}.
\ee
Given the metric in the form,
\be\label{metanz}
ds^2 = - e^{-4A(r)}f(r) dt^2 + e^{2B(r)} \lb {dr^2 \over f(r)} + {r^2 
\over b^2} (d\theta^2 + d \vec x^2 ) \rb,
\ee
the action \ref{onshellacn} becomes,
\be\label{ba1}
I_B^{bulk} = {\beta V \eta \over 8 \pi G_5} \lb {\tilde R^2 f(\tilde R) 
\over b^3} + 
{\tilde R^3 f(\tilde R) A'(\tilde R) \over b^3} \rb.
\ee 
 We have to add the following GH term to the action \ref{gauge-sugra-acn},
\be\label{gh}
I_B^{GH} = {1 \over 8 \pi G_5} \int d^4 x \sqrt{-h} \Theta,
\ee
where $h$ is induced metric on the boundary, $\Theta = - \nabla_{\mu} n^{\mu}$
 and $n$ is unit normal in the $r$ direction, $n_r = \sqrt{g_{rr}}$. Using 
metric \ref{metanz}, the GH term becomes,
\be\label{gh2}
I_B^{GH} = - {\beta V \eta \over 8 \pi G_5} \lb {\tilde R^2 (\tilde R 
f'(\tilde R)+ 2f(\tilde R)(3 + \tilde R 
A'(\tilde R))) \over 2 b^3} \rb.
\ee
Hence the on-shell black hole action is given by,
\be\label{ba2}
I_B=  - {\beta V \eta \over 8 \pi G_5} \lb {\tilde R^2 (4 f(\tilde R) 
+ \tilde R f'(\tilde R)) \over 2 
b^3} \rb.
\ee
The function $f(r)$ is given by,
\be
f(r) = -{M \over r^2} + {r^2 \over b^2} H_1 H_2 H_3
\ee and
\be
H_I = 1+ {q_I \over r^2}.
\ee
Putting $f(r)$ and $H(r)$ in equation \ref{ba2} we get (in $b=1$ unit),
\be\label{ba3} 
I_B = -{\beta V \eta \over 8 \pi G_5} \left [- M + (q_1 q_2 + q_2 q_3 
+ q_3 q_1) + 2 \tilde R^2 (q_1 + q_2 + q_3) + 3 \tilde R^4 
\right ].
\ee

Now we will compute the soliton action on equation of motion following 
the same steps as above. Soliton metric is given by \ref{solmetiso}. 
Using that, the bulk soliton action is given by,
\be\label{sola1}
I_{S}^{bulk} = {\beta_S V \eta_S \over 8 \pi G_5} \left[{\tilde R^4}
\lb (H_1 H_2 H_3)^{2/3} - {\mu^4 \over \tilde R^4}\rb \right]
\ee
and the GH boundary action is given by,
\be\label{sola2}
I_S^{GH} = -{\beta_S V \eta_S \over 8 \pi G_5}\left [{4 \tilde R^4}
(H_1 H_2 H_3)^{2/3}\lb
1 - {\mu^4 \over 2 \tilde R^4 (H_1 H_2 H_3)^{2/3}}\rb \right].
\ee 
Hence the on-shell soliton action is given by,
\ben\label{solacn1}
I_S &=& I_S^{bulk}+I_S^{GH} \nonumber \\
&=& {\beta_S V \eta_s \over 8 \pi G_5} \lb {-3 \tilde R^4(H_1 H_2 H_3)^{2/3}
 + \mu^4 } \rb.
\een
Using the relation between $(\beta, \eta)$ and $(\beta_S, \eta_S)$ we 
find the subtracted action is,
\ben
I&=& - {\beta V \eta_S \over 16 \pi G_5} \lb M  + {2 \over 3 }(q_1^2 + q_2^2 + q_3^2) - {2\over 3}(q_1 q_2 + q_2 q_3 + q_3 q_1)-\mu^4  \rb. 
\een


\end{document}